\documentclass[useAMS,usenatbib]{mn2e}

\usepackage{epsfig}
\usepackage{amssymb}
\newcommand{\mdot}{M$_{\odot}$}

\usepackage{mathptmx}

\title[Jet-driven outflows of cold gas in 3C293]{The location and impact of jet-driven outflows of cold gas: the case of 3C293}
\author[E. K. Mahony et al.]{E. K. Mahony$^{1}$\thanks{E-mail:mahony@astron.nl}, R. Morganti$^{1,2}$, B. H. C. Emonts$^{3,4}$, T. A. Oosterloo$^{1,2}$ and \and C. Tadhunter$^{5}$ \\
$^{1}$ASTRON, the Netherlands Institute for Radio Astronomy, Postbus 2, 7990 AA, Dwingeloo, The Netherlands.\\
$^{2}$Kapteyn Astronomical Institute, University of Groningen, Postbus 800, 9700 AV Groningen, The Netherlands.\\
$^{3}$Centro de Astrobiolog\'ia (INTA-CSIC), Ctra de Torrej\'on a Ajalvir, km 4, 28850 Torrej\'on de Ardoz, Madrid, Spain \\
$^{4}$Australia Telescope National Facility, CSIRO Astronomy and Space Science, PO Box 76, Epping, NSW 1710, Australia.\\
$^{5}$Department of Physics \& Astronomy, University of Sheffield, Sheffield S3 7RH. 
}

\begin{document}

\date{Accepted 2013 .... Received 2013 ...; in original form 2013 ...}

\pagerange{\pageref{firstpage}--\pageref{lastpage}} \pubyear{2013}

\maketitle

\label{firstpage}

\begin{abstract}

The nearby radio galaxy 3C293 is one of a small group of objects where extreme outflows of neutral hydrogen have been detected. However, due to the limited spatial resolution of previous observations, the exact location of the outflow was not able to be determined. In this letter, we present new higher resolution VLA observations of the central regions of this radio source and detect a fast outflow of HI with a FWZI velocity of $\Delta v\sim$1200\,km\,s$^{-1}$ associated with the inner radio jet, approximately 0.5\,kpc west of the central core. We investigate possible mechanisms which could produce the observed HI outflow and conclude that it is driven by the radio-jet. However, this outflow of neutral hydrogen is located on the opposite side of the nucleus to the outflow of ionised gas previously detected in this object. 

We calculate a mass outflow rate in the range of $8-50$\,\mdot\,yr$^{-1}$ corresponding to a kinetic energy power injected back into the ISM of $1.38 \times 10^{42} - 1.00 \times 10^{43}$ erg\,s$^{-1}$ or $0.01 - 0.08$\,per\,cent of the Eddington luminosity. This places it just outside the range required by some galaxy evolution simulations for negative feedback from the AGN to be effective in halting star-formation within the galaxy. 

\end{abstract}


\section{Introduction}

The tight correlations observed between the mass of the supermassive black hole and the mass and stellar velocities of the bulge provide compelling evidence that the evolution of galaxies and their central black holes are strongly linked \citep{Kormendy1995, Magorrian1998, msigma}. This is generally attributed to feedback mechanisms, where the interplay between the supermassive black hole and surrounding medium regulates the growth of the galaxy \citep{croton06,bower06}. This feedback often takes the form of large outflows of gas, quenching star formation in the host galaxy and halting accretion onto the AGN. While the required outflows can be modelled in galaxy formation simulations, the observational evidence is limited, particularly on scales close to the black hole.

There is a range of possible physical mechanisms thought to be responsible for driving these outflows, the most widely accepted being supernovae driven winds from starbursts \citep{Heckman1990} or powerful winds driven from the nucleus itself \citep{Silk1998}. However, there is increasing evidence that in radio loud AGN, the interaction between the radio jets and the ISM of the host galaxy can also play a major role \citep{Nesvadba2008, Holt2008, Guillard2012}. Massive and fast outflows of HI and CO, possibly driven by this interaction, have been observed in a growing number of objects (see e.g. \citealt{Morganti2005a, Dasyra2012, Feruglio2010}). Tracing these outflows with observations at high spatial resolution allows us to quantify the impact of these AGN on the evolution of the host galaxy, providing constraints for numerical simulations. Pinpointing the location of these outflows also enables us to derive crucial parameters, such as the mass outflow rates and kinetic energy involved, which we can compare to predictions from galaxy evolution simulations. 

Previous observations of the nearby radio galaxy 3C293 have detected fast outflows of neutral hydrogen using the Westerbork Synthesis Radio Telescope (WSRT, \citealt{Morganti2003}). However, the spatial resolution was not high enough to conclusively determine where the outflows were located. In a follow-up study, outflows of ionised gas were observed in this object using long-slit spectroscopy of the optical emission lines \citep{Emonts2005}. These observations show that the ionised outflows originate from the eastern radio jet, approximately 1\,kpc from the nucleus. Due to the similarity of the HI and ionised gas profiles, the authors concluded that the HI outflow was also associated with the eastern radio jet. 

In this letter we present higher resolution data from the Karl G. Jansky Very Large Array (VLA) where we detect fast outflows of neutral hydrogen associated with the western inner jet approximately 0.5 kpc from the central core. 

\section{Observations and data reduction}

\subsection{Previous observations}

3C293 is a nearby radio galaxy associated with the optical source UGC8782 at a redshift of $z=0.045$ \citep{Sandage}. On subarcsec scales the compact steep spectrum core is resolved into multiple knots along the radio jet \citep{Beswick2004}. This restarted radio emission, along with the disturbed optical morphology \citep{Heckman1986}, suggest that this system has recently undergone merger activity.

Neutral hydrogen was first detected in this system by \citet{Baan1981} who discovered 480\,km\,s$^{-1}$ absorption associated with the rotating disk. This was followed by a more detailed study of the HI associated with the dust lanes on VLBI scales \citep{Beswick2002, Beswick2004}. Extremely broad absorption in 3C293 was then detected using the Westerbork Synthesis Radio Telescope (WSRT). These observations used a bandwidth of 20\,MHz and velocity resolution of 4\,km\,s$^{-1}$ allowing for large velocity coverage \citep{Morganti2003}. The broad absorption component was detected towards the core of the radio galaxy and has a FWZI velocity of 1400\,km\,s$^{-1}$, of which $\sim 1000$\,km\,s$^{-1}$ is blueshifted compared to the systemic velocity. However, the $\sim$10\,arcsec spatial resolution of WSRT meant that it could not be distinguished if this fast outflow was coming from the central core or from the inner radio lobes. 

\subsection{New observations}

Observations were carried out with the VLA in the A-array to obtain a spatial resolution of $\sim$1\,arcsec. A single subband of 256 channels was used with 14\,km\,s$^{-1}$ velocity resolution and a total bandwidth of 16\,MHz. The band was centred at 1359.13\,MHz, roughly corresponding to the central frequency of the broad absorption that was previously detected with WSRT. 

The source was observed for 100 mins on August 18 2011 with 3C286 used for phase, bandpass and absolute flux calibration. To ensure that a good bandpass calibration was obtained 3C286 was observed every 20 mins. The data was reduced and hanning-smoothed using CASA. For the continuum subtraction, care was taken to exclude channels which were close to the broad absorption feature to avoid introducing any artifacts that may affect the analysis. To achieve the optimal spatial resolution, images were made with uniform weighting resulting in a resolution of 1.2$\times$1.3 arcsec. The rms reached was 0.8\,mJy/beam/channel (after hanning-smoothing).

\section{Results \& Discussion}

\begin{figure}
\includegraphics[width=\linewidth, trim=120 20 120 120, clip]{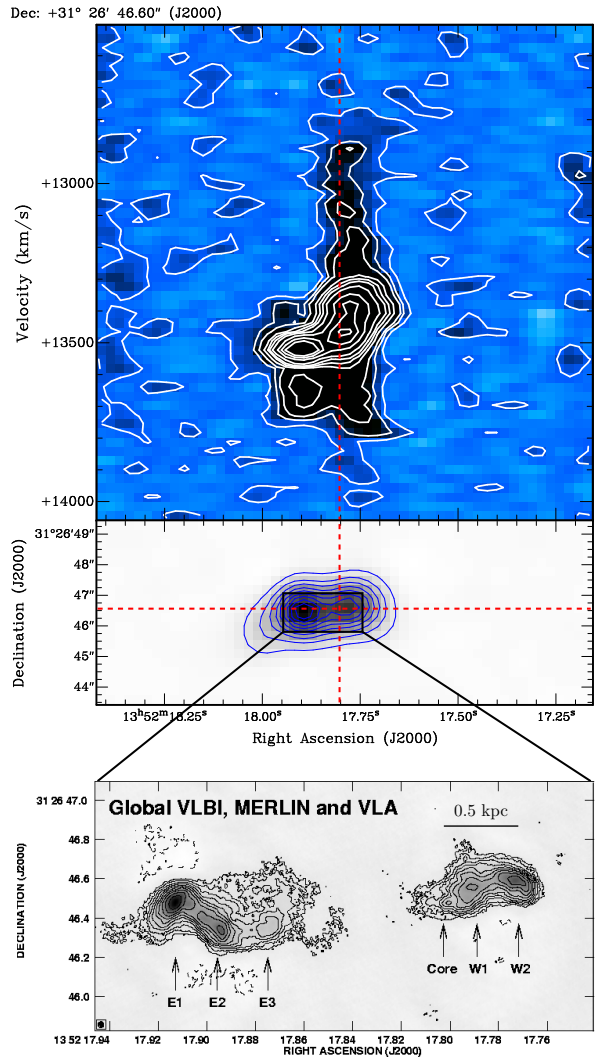}
\caption{{\bf Top:} Position-velocity diagram extracted along the radio axis of the continuum source. {\bf Middle:} The continuum image from the new VLA observations. The horizontal red line marks the axis along which the position-velocity diagram was extracted and the vertical red line indicates the position of the core taken from the VLBI image below. {\bf Bottom:} Combined 1.4\,GHz global VLBI, MERLIN and VLA image of the central regions of 3C293 from \citet{Beswick2004}. \label{fig1}}
\end{figure} 

The middle panel of Figure \ref{fig1} shows the VLA continuum image in which the inner radio lobes of 3C293 can be seen. The VLBI image of \citet{Beswick2004} is shown underneath for comparison. The top panel shows the position-velocity diagram extracted along the radio axis. An extremely broad component ($\Delta v>$1000\,km\,s$^{-1}$) is immediately clear against the western side of the radio source. 

\begin{figure}
\centering{\epsfig{file=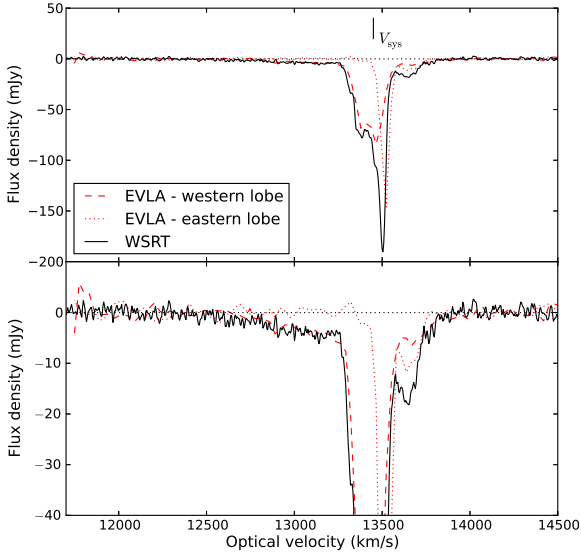, width=\linewidth}}
\caption{VLA spectral profiles from each radio lobe (red dashed lines) compared to the WSRT spectral profile (solid black line). The broad component seen in the WSRT observations is fully recovered in the VLA observations, but only observed in front of the western radio lobe. The systemic velocity ($V_{\mathrm{sys}}$=13450\,km\,s$^{-1}$) of this galaxy is also marked. The bottom panel shows the same data on a smaller vertical scale to more clearly show the broad absorption. \label{overlay}}
\end{figure}

To verify that this broad absorption is only observed in front of one of the radio lobes, Figure \ref{overlay} shows the HI profiles from each of the eastern and western radio lobes. The HI profile from WSRT observations is overlaid for comparison. Strong narrower absorption lines from the rotating disk can be seen across the radio source, but the very broad component (with a FWZI velocity $\Delta v\sim$1200\,km\,s$^{-1}$) is only observed in front of the western radio lobe. Furthermore, this western profile is very similar to the WSRT profile, indicating that all of the broad absorption detected in the original WSRT observations has been recovered at higher spatial resolution. This indicates that the observed outflow is localised to a region smaller than $\sim$1.2 arcsec, or 1\,kpc at this redshift (assuming $H_0=71$ km\,s$^{-1}$, $\Omega_0=0.27$, $\Omega_\Lambda=0.73$).

While this VLA image represents a significant increase in spatial resolution over the previous WSRT image, it is still not sufficient to completely separate the fainter core from the western radio lobe (see Figure \ref{fig1}). Nevertheless, the absorption appears clearly offset toward the western side and offset compared to the location of the core (marked by the vertical red, dashed line). Furthermore, the core has a much lower continuum flux density than the lobe, just 27.1\,mJy at 1.4 GHz \citep{Beswick2004}, thus to observe this broad absorption feature against that continuum source would require a remarkably high peak optical depth of $\tau$=0.26 or an integrated optical depth of $\int\tau dv \approx \tau_{\mathrm{peak}} \times \Delta \nu (\mathrm{FWHM})=186$. As such, we conclude that the broad absorption is observed in front of the western lobe which has a continuum flux of 1.17\,Jy. This leads to a peak optical depth of $\tau$=0.006 and an integrated optical depth of $\int \tau dv \approx 4.3$, more consistent with optical depths measured in other cases of HI outflows \citep{Morganti2005a}. 

By comparing with the VLBI image we deduce that the very broad absorption is located in front of the western lobe up to 0.5 kpc from the core. This suggests that the radio-jet is responsible for driving the observed outflow of neutral hydrogen as previously speculated by \citet{Morganti2003} and \citet{Emonts2005}. 

However, before we can conclusively say that the radio jet is driving the outflow, we need to address the question of whether the jet in 3C293 is capable of driving such a high-velocity outflow. Using hydrodynamical simulations, \citet{Wagner2012} show that the radio jet is capable of accelerating clouds in the ISM to high velocities if the ratio of the jet power to Eddington luminosity $\eta=P_{\mathrm{jet}}/L_{\mathrm{Edd}}$ is above some critical value $\eta_{\mathrm{crit}}$. In most cases $\eta_{\mathrm{crit}}\gtrsim10^{-4}$ \citep{Wagner2011}, but can be as high as $\eta_{\mathrm{crit}}=10^{-2}-10^{-1}$ for an ISM with large cloud complexes. In the case of 3C293 $\eta=0.4$ (given $P_{\mathrm{jet}}=5.1\times10^{45}$\,ergs\,s$^{-1}$; \citealt{Wagner2012}) meaning that the radio jet is indeed capable of driving the observed outflow. A jet-driven outflow is also consistent with the highly (super-thermally) excited CO emission observed in 3C293, believed to be induced by shocks created by this jet-ISM interaction \citep{Papadopoulos2008, Papadopoulos2010}. 

Additionally, as shown in \citet{Emonts2005} the nucleus of 3C293 is too weak to drive an outflow of this magnitude due to radiation pressure, particularly out to 0.5 kpc from the core. Similarly, the starformation rate in this galaxy of $\lesssim$ 4\,\mdot\,yr$^{-1}$ \citep{Papadopoulos2010} is much less than seen in other objects exhibiting starburst driven winds (see e.g. \citealt{Veilleux2005} and references therein). 

By fitting multiple gaussian components to the HI profile we find that the broad component has a FWHM of $v=719\pm129$\,km\,s$^{-1}$ centred at 13428\,km\,s$^{-1}$. While this is extremely broad, it is only slightly blueshifted from the systemic velocity of 13450\,km\,s$^{-1}$. This could imply a slightly different scenario in which (part of) the broad HI gas is highly turbulent gas within the galaxy's disk that is in rotation around the core. In that case, the disk's rotation at the location of the inner western lobe \citep{Emonts2005} would mimic a net blueshift of order $\sim 50-100$\,km\,s$^{-1}$ of the HI gas, in agreement with the central velocity of the broad HI absorption. This scenario could be similar to that of the intermediate HI absorption component described by \citet{Beswick2004} against the inner western radio lobe. 

However, even in this scenario, the very large velocity dispersion of the gas would most likely be induced by the propagating radio jet (similar to the jet-induced turbulence of molecular gas found in the center of 3C293; \citealt{Guillard2012}). Moreover, a jet-ISM interaction that induces such an extreme turbulence would be highly disruptive to the disk and the jet would probably be dragging gas in outflow up from regions much closer to the nucleus (see e.g. \citealt{Sutherland2007}).

\subsection{Other cases of jet-driven outflows}

The idea that the radio jet can affect the surrounding ISM is not new. There have been many studies showing alignments between the radio jet and optical structures in the NLR in Seyfert galaxies \citep{Whittle1988, Capetti1996, Whittle2004, Rosario2010}. Likewise many radio galaxies also show alignment of the extended emission line region along the jet axis \citep{McCarthy1987, Best1998, Tadhunter2000}. The fact that this jet-ISM interaction can also produce outflows of cold, neutral gas has been seen in other objects such as IC5063 \citep{Morganti2007}, 3C305 \citep{Morganti2005} and 4C12.50 (\citealt{Holt2010}, Morganti et al., submitted). In many of these cases, outflows with similiar kinematics have also been detected in ionised \citep{Holt2008, Holt2010}, and molecular gas \citep{Morganti2013, Guillard2012, Dasyra2012}. Outflows of ionised gas have been detected in 3C293, with the main outflow occuring at the location of the eastern inner hot-spot \citep{Emonts2005}. Thus, unlike other cases of jet-induced outflows which have been observed at similar or higher resolution (i.e. IC5063 and 3C305), the main ionised gas outflow observed in 3C293 is not located in the same region as the HI outflow.   

\subsection{Comparison with the ionised gas outflows} \label{bjorn}

\begin{figure}
\centering{\epsfig{file=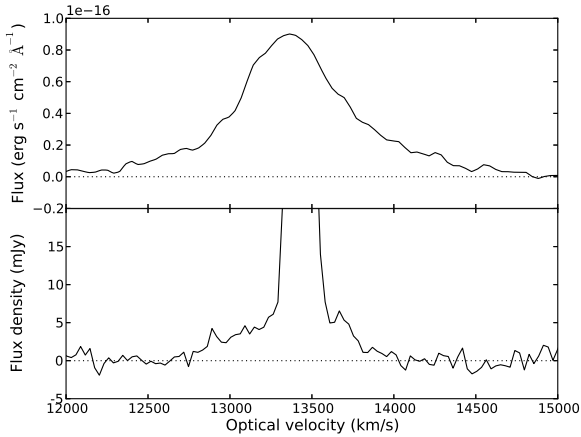, width=\linewidth}}
\caption{The top panel shows the $\lambda$6719 [SII] emission line from the `C' region presented in \citet{Emonts2005} which is compared to the inverted HI profile obtained from the VLA observations shown in the bottom panel.\label{ionisedoverlay}}
\end{figure}

It is important to note that while \citet{Emonts2005} also conclude that the outflow is driven by the radio jet, this was based on the similarity of the [OII] profile and the WSRT HI profile and, as such, it was assumed that both the ionised gas and HI outflow were associated with the eastern jet. The new VLA data presented here show that the HI outflow is in fact located in front of the western radio jet. Looking more closely at the previous observations of the optical emission lines of \citet{Emonts2005}, it is clear that there is also a broad component in the central region (denoted `C' in \citealt{Emonts2005}), which encompases the western inner radio jet where we detect the HI outflow. Figure \ref{ionisedoverlay} shows the broad component of the $\lambda$6719 [SII] emission line from \citet{Emonts2005} in the top panel and the (inverted) HI absorption detected in the VLA observations in the bottom panel for comparison. This shows that the [SII] emission line spans an even larger velocity range than the HI absorption, indicating that outflowing ionised gas could also be present at the location of the outflowing neutral gas.

While the kinematics of the neutral and ionised gas outflows broadly agree at this particular location in the galaxy, this does not explain why the most extreme ionised gas outflows are observed in the eastern jet while the HI observations show no hint of any outflows in this region. This difference could be partially explained by observational biases; since the HI is observed in absorption we only detect the gas in front of a strong continuum source while in the optical we are limited by any obscuration. In addition, since we are probing different phases of the gas there is not necessarily any reason why they should be co-located. Furthermore, it has also been shown by \citet{Gaibler2011} that jets can evolve very asymmetrically in an inhomogeneous ISM. Follow-up observations to map the ionised gas outflows across the galaxy are the subject of a future paper.

\subsection{Impact of the jet}

Having determined that the fast outflow of HI is most likely driven by the radio jet, we can calculate the outflow rate using the following formula \citep{Heckman2002}:

\begin{equation}
\dot{M} = 30 \frac{\Omega}{4\pi} \frac{r_*}{1\,\mathrm{kpc}} \frac{N_{\mathrm{HI}}}{10^{21}\mathrm{cm}^{-2}} \frac{v}{300\,\mathrm{km\,s}^{-1}}  $\mdot$\,{\rm yr}^{-1}
\end{equation}

\noindent where $\Omega$ is the solid angle subtended by the outflow (assumed to be $\pi$), $r_*$ the radius, $v$ the velocity of the outflow and N$_{\mathrm{HI}}$ is the column density. The shallow, broad component of the HI profile has an optical depth of $\tau$=0.006 corresponding to a column density of $6.6\times10^{21}$\,cm$^{-2}$. Due to the extreme conditions close to the centre of an AGN we assume a T$_{\mathrm{spin}}$=1000\,K \citep{Holt2006, Bahcall+Ekers}. From this column density and assuming a spherical geometry of the outflowing gas such that the HI column observed also extends 0.5\,kpc, we derive a density of $\sim$4.3\,cm$^{-3}$. However, if the outflowing region is smaller, or more clumpy, the density could be much higher. Due to the spatial resolution of these observations, it is unclear whether the outflow extends along the entire jet or is located just at the hotspot. VLBI observations are required to constrain the size of the outflow further, but for now we assume a radius of r$_*=0.5$\,kpc corresponding to the distance from the core to the hotspot.

With our current data it is difficult to estimate the velocity of the outflowing gas due to the fact that it is impossible to distinguish what fraction of the broad HI component is outflowing and what fraction is turbulent motions within the disk. As such, we use a range of velocities from a fairly conservative estimate of the outflowing velocity $v=100$\,km\,s$^{-1}$ (in which case the broad absorption is dominated by turbulent motions of the gas) to $v=600$\,km\,s$^{-1}$ assuming that all the gas in the broad feature is outflowing (here the velocity is taken as the FWZI/2). This results in a mass outflow rate in the range of $8 - 50$\,\mdot\,yr$^{-1}$. 

From this mass outflow rate we can calculate the energy loss rate following \citet{Holt2006}:

\begin{equation}
\dot{E} = 6.34 \times10^{35} \frac{\dot{M}}{2} \left(v^2 + \frac{{\rm FWHM}^2}{1.85}\right) {\rm erg}\,{\rm s}^{-1} 
\end{equation}

\noindent Using the same parameters as before: $\Omega=\pi$, $r_*=0.5$\,kpc, $N_{\mathrm{HI}}=6.6\times10^{21}$cm$^{-2}$, $v=100 - 600$\,km\,s$^{-1}$ and a covering factor C$_f=1$, we calculate an energy loss rate in the range $1.38 \times 10^{42} < \dot{E} < 1.00 \times 10^{43}$ erg\,s$^{-1}$. Using a black hole mass of 10$^8$\mdot and Eddington ratio of 0.017 given in \citet{Wu2009}, this energy loss rate corresponds to $0.01-0.08$\,per\,cent of the Eddington luminosity ($L_{\mathrm{Edd}}$). 

Many galaxy evolution simulations require outflows of approximately 5$-$10\,per\,cent of $L_{\mathrm{Edd}}$ to produce the observed M-$\sigma$ relation (see e.g. \citealt{DiMatteo2005, Booth2009}). However, the two-phase ISM model of \citet{Hopkins2010} requires an order of magnitude less energy to be injected back into the ISM (approximately 0.5\,per\,cent). Using this two-phase model, the observed outflow of HI is close to the level needed to clear the gas out of the galaxy, even before taking into account outflows in other gas phases.

\section{Conclusions and future work}

This letter presents new VLA data of the nearby radio galaxy 3C293 where we detect a fast outflow of neutral hydrogen in front of the western radio jet, approximately 0.5\,kpc from the central nucleus. We conclude that the radio jet is the most likely candidate for driving this outflow. The HI outflow has a FWZI velocity of $\Delta v\sim$1200\,km\,s$^{-1}$ and a mass outflow rate in the range of $8-50$\,\mdot\,yr$^{-1}$. The kinetic energy power injected back into the ISM is in the range of $1.38 \times 10^{42} - 1.00 \times 10^{43}$ erg\,s$^{-1}$ corresponding to $0.01 - 0.08$ \,per\,cent of the Eddington luminosity. This places it just below the 0.5\,per\,cent required by the galaxy evolution simulations of \citet{Hopkins2010} to heat or drive-out the gas, thereby halting star formation in the host galaxy. 

Intriguingly, the HI outflow is associated with the opposite radio jet to the fast outflow of ionised gas previously detected by \citet{Emonts2005}. Investigating this further revealed that the kinematics of the ionised gas at the location of the western jet is consistent with the neutral gas kinematics, but more work is needed to properly characterise the dynamics and kinematics of the ionised gas in this system. 
 
These higher resolution observations of 3C293 adds another case to the few examples of radio sources where outflows of neutral hydrogen are offset from the nucleus on scales of hundreds of parsecs. By resolving the location of the outflowing gas and comparing the characteristics of the different phases of the outflowing gas (ionised, neutral hydrogen and molecular) we can provide constraints for understanding the ongoing process of interaction, heating of the gas and subsequent cooling.

\section*{Acknowledgements}

The authors wish to thank Emmanuel Momjian for useful discussions and the anonymous referee for several suggestions which improved the paper. The National Radio Astronomy Observatory is a facility of the National Science Foundation operated under cooperative agreement by Associated Universities, Inc.

\bibliographystyle{mn2e}
\setlength{\bibhang}{2.0em}
\setlength\labelwidth{0.0em}

\bibliography{3c293}

\bsp

\label{lastpage}

\end{document}